\documentclass[conference]{IEEEtran}

\usepackage{amsfonts}
\usepackage{algorithm,caption}
\usepackage{graphicx,cite,calc}
\usepackage{color,amsthm}
\usepackage{amsmath}
\usepackage{algorithm}
\usepackage{algpseudocode}
\usepackage{url}
\newtheorem{thm}{Theorem}
\newtheorem{lem}{Lemma}

\newtheorem{examp}{Example}

\newtheorem{myremark}{Remark}

\begin{document}

\title{Multi-Antenna Coded  Caching}
 
\author{ Seyed Pooya Shariatpanahi$^1$, Giuseppe Caire$^{2}$, Babak Hossein Khalaj$^{3}$ \\[4mm] 

1: School of Computer Science, Institute for Research in Fundamental Sciences (IPM), Tehran, Iran. \\
2: Technical University of Berlin, 10587 Berlin, Germany. \\
3: Department of Electrical Engineering,
Sharif University of Technology, Tehran, Iran.\\
(emails: pooya@ipm.ir, caire@tu-berlin.de, khalaj@sharif.edu)\\ } 
 
\maketitle

\begin{abstract}
In this paper we consider a single-cell downlink scenario where a multiple-antenna base station delivers contents to multiple cache-enabled user terminals. Based on the multicasting opportunities provided by the so-called \emph{Coded Caching} technique, we investigate three delivery approaches. Our baseline scheme employs the coded caching technique on top of max-min fair multicasting. The second one consists of a joint design of Zero-Forcing (ZF) and coded caching, where the coded chunks are formed in the signal domain (complex field). The third scheme is similar to the second one with the difference that the coded chunks are formed in the data domain (finite field). We derive closed-form rate expressions where our results suggest that the latter two schemes surpass the first one in terms of Degrees of Freedom (DoF). However, at the intermediate SNR regime forming coded chunks in the signal domain results in power loss, and will deteriorate throughput of the second scheme. The main message of our paper is that the schemes performing well in terms of DoF may not be directly appropriate for intermediate SNR regimes, and modified schemes should be employed.
\end{abstract}

\section{Introduction}

Caching content near end-users is one of the most promising solutions proposed for next generation wireless networks such as 5G \cite{Bastug_2014}. The main reason of the effectiveness of this approach is that a significant portion of mobile traffic is  multimedia-like which makes demands predictable \cite{UMTS_Forum}. On the other hand, memory hardware is much cheaper than bandwidth, and is abundantly available at mobile devices. Thus, using a technique which turns memory into bandwidth (i.e., \emph{Caching}) is of high interest.

The surprising finding of \cite{Maddah-Ali_Fundamental_2014} shows that by careful caching at end-users the delivery bandwidth burden can be greatly relaxed by multicasting coded file chunks to these users. The main idea of this so-called \emph{Coded Caching} approach is to cache non-identical file chunks among different users, which will provide coded multicasting opportunities. This idea has been extended to other scenarios such as hierarchical caching \cite{Karamchandi_2016}, D2D networks \cite{Ji_2016}, and multi-server setup \cite{Pooya_2016}.

Since the coded caching scheme is not originally designed for wireless scenarios (such as 5G) this approach should carefully be adapted to specific characteristics of wireless channels. In order to achieve this goal we consider a multi-antenna transmitter (e.g., a base station) delivering contents to multiple single-antenna receivers (e.g., user terminals), i.e., cache-enabled MISO broadcast. Since the coded caching approach is designed to provide multicasting opportunities, the baseline scheme we consider is using the multiple antennas at the transmitter to design max-min fair multicast transmissions to receiver groups. The second approach we consider is employing the transmit antennas' multiplexing gain  to enlarge the size of multicast groups, following ideas in \cite{Pooya_2016}. We investigate this approach by assuming coded file chunks formed in both the signal domain (complex field) and the data domain (finite field). For all the aforementioned schemes we derive delivery rate at finite SNR.

It should be noticed that 
a conventional approach using caching in the non-cooperative traditional way would simply cache a fraction of the library at each user, and serve a number of users (smaller than the number of transmit antennas) via spatial multiplexing and single-user codes. In contrast to our proposed schemes, such solution does not benefit from the cooperative caching gain  making it non-scalable for large networks.

While there are other works which have considered coded caching in wireless scenarios such as \cite{Jingjing, Navid, MAT, Hachem} our paper addresses the problem at finite SNR. Such finite SNR analysis proposes important design guidelines not revealed in earlier (high SNR) DoF analysis of wireless coded caching papers. Moreover, although the paper \cite{Ngo_2016} considers a finite SNR setup, in contrast to our paper, their scheme is designed for a massive MIMO scenario with more transmit antennas than users. In this case, if full Channel State Information at the Transmitter (CSIT) is available, there is no need for multicasting opportunities provided by coded caching.

The rest of the paper is organized as follows. In Section \ref{Sec:Model} we describe the model. In Section \ref{Sec:Max-Min} we describe the baseline scheme of Max-Min Fair Multicasting. In Sections \ref{Sec:ComplexField}  and  \ref{Sec:FiniteField} we analyze Multi-Antenna Coded Caching. Section \ref{Sec:Compare} compares the three schemes, and finally, Section \ref{Sec:Conclusions} concludes the paper.

In this paper we use the following notations. We use $(.)^H$ to denote the Hermitian of a complex matrix. Let $\mathbb{C}$ and $\mathbb{N}$ denote the set of complex and natural numbers and $||.||$ be the norm of a complex vector. Also $[1:m]$ denotes the set of integer numbers $\{1,\dots,m\}$, and $\oplus$ represents addition in the corresponding finite field. For any vector $\mathbf{v}$, we define $\mathbf{v}^{\perp}$ such that $\mathbf{v}^H\mathbf{v}^{\perp}=0$.

\section{Model}\label{Sec:Model}
We consider downlink transmission from a multiple-antenna base station (BS) with $L$ antennas having access to a library of $N$ files
$\{W_1, \ldots, W_N\}$, each of size $F$ bits. The transmission is to $K$ single antenna User Terminals (UT). The wireless channel from the BS to the UTs is represented by the matrix $\mathbf{H}^H \in \mathbb{C}^{K \times L}$ 
(complex baseband discrete-time channel model). In addition, we consider $\mathbf{H}$ to be constant over large blocks of $B \gg 1$ channel uses in the time-frequency domain. Also, we assume full CSIT is available, and $K \geq L$.

Let us represent data by $m$-bit symbols in the finite field $\mathbb{F}_{2^m}$. Consider a one-to-one map $\psi$ from $\mathbb{F}_{2^m}$ to $n$ complex numbers belonging to a Gaussian codebook $\mathcal{C}$, i.e., 
$
\psi : \mathbb{F}_{2^m} \rightarrow \mathbb{C}^{n}
$,
which constitute the transmit signal from BS antennas passed through $n$ channel uses. We also assume a power constraint on the codebook as follows: If $x$ is an $m$-bit symbol, then we should have $\mathbb{E}\left[|\psi(x)|^2\right] \leq n \times SNR$.
The operator $\psi$ encodes a vector of symbols element-wise. For ease of presentation we denote the encoded version of file $W$ with $\psi(W)=\tilde{W}$ throughout the paper.

Consider $n$ channel uses. Then, the received signal at user $k$ is given by 
\begin{equation} \label{Eq_Channel_Model}
\underline{\mathbf{y}}_k = \mathbf{h}_k^H \underline{\mathbf{X}} + \underline{\mathbf{z}}_k, 
\end{equation}
where $\underline{\mathbf{y}}_k, \underline{\mathbf{z}}_k \in \mathbb{C}^{1 \times n}$ denote the received signal sequence (over $n$ channel uses) at receiver $k$ and the corresponding
additive white Gaussian noise sequence, with i.i.d. components $\sim\mathcal{CN}(0,1)$, $\mathbf{h}_k$ is the $k$-th column of $\mathbf{H}$, and $\underline{\mathbf{X}} \in \mathbb{C}^{L \times n}$ is the 
space-time block of transmitted coded signal collectively transmitted by the BS over the $n$ channel uses. Also, we consider the total transmit power constraint 
\begin{equation}\label{Eq_Transmit_Power_Const}
\frac{1}{nL}\sum_{i,j}\mathbb{E}\left[|\underline{\mathbf{X}}_{i,j}|^2\right] \leq SNR.
\end{equation}

In the system at hand, the users are equipped with a cache memory of capacity $MF$ bits. Before the network operations (e.g., at off-peak times, or at home, downloading from 
a home INTERNET access) each user $k$ has stored in its cache a message $Z_k = Z_k(W_1, \ldots, W_N)$, where $Z_k(\cdot)$ denotes a function of the library files with entropy
not larger than $MF$ bits. This operation is referred to as the {\em cache content placement}, and it is performed once and at no cost. During the network operation, 
users place requests for files in the library. We let $d_k \in [1:N]$ denote the request of user $k$ and $\mathbf{d} = (d_1, \ldots, d_K)$ be the request vector. 

Upon a set of requests $\mathbf{d}$ at the \emph{content delivery} phase, the BS transmits a coded signal, such that at the end of transmission all users can reliably decode
their requested files. Notice that user $k$ decoder, in order to produce the decoded file $\widehat{W}_{d_k}$, makes use of its own cache content $Z_k$ as well as of its own received signal from the wireless channel. 

In this work we focus on the worst-case (over the users) delivery rate at which the system can serve any users requesting any file of the library. Consider the MIMO channel at hand, and suppose that the coded multicasting codeword
is formed by the concatenation of subcodewords $U_S$, where each subcodeword is dedicated to a subset $S \subseteq [1:K]$ of users, of length $\mathcal{L}(U_S) F$ bits each. Let $C(SNR,S,\mathbf{H})$ (in bit/s) denote the multicast rate at which the BS can communicate a common message to all users in subset $S$. It follows that the total transmission time necessary to deliver all multicast subcodewords is given by 
\begin{equation}\nonumber
T = \sum_{S \subset [1:K]} \frac{\mathcal{L}(U_S) F}{C(SNR, S, \mathbf{H})}. 
\end{equation} 

Since time $T$ is necessary for each user to be able to decode its own request file of $F$ bits, the system \emph{symmetric rate} can be defined as the ``goodput'' (useful bits per second) 
at which each user is served. Since each user is able to decode a file of $F$ bits after time $T$, the per user symmetric rate is given by 
\begin{equation} \label{Eq_SymRate_Model}
R_{\mathrm{sym}} = \frac{F}{T} = \left [  \sum_{S \subset [1:K]} \frac{\mathcal{L}(U_S)}{C(SNR, S, \mathbf{H})} \right ]^{-1}. 
\end{equation}

\section{Coded Caching with Max-Min Fair Multicasting}\label{Sec:Max-Min}

In this case, the cache content placement works exactly as in Maddah-Ali and Niesen scheme \cite{Maddah-Ali_Fundamental_2014}. For the case of $t = \frac{MK}{N} \in \mathbb{N}$, each file is partitioned into
${K \choose t}$ non-overlapping subfiles as:
\begin{equation}\nonumber
W_n = \{W_{n,\tau} : \tau \subset [1:K], |\tau| = t\}, \;\;\;\; \forall \;\; n \in [1:N].
\end{equation}
Each user $k$ stores in its cache all subfiles such that $k \in \tau$. These are ${K - 1 \choose t - 1}$, such that the cache memory is completely used
since $\frac{NF}{{K \choose t}} {K - 1 \choose t - 1} = MF$.

In the content delivery phase, let $\mathbf{d} = (d_1, \ldots, d_K)$ be the current demand vector. For all subsets $S \subseteq [1:K]$ of size $|S| = t + 1$ (denoted in the following as
$(t+1)$-subsets), the BS forms the coded message
\begin{equation} \label{Eq_Xor-CodedCache}
U_S = \oplus_{k\in S} W_{d_k, S\setminus \{k\}}. 
\end{equation}

In Maddah-Ali and Niesen proposal \cite{Maddah-Ali_Fundamental_2014} the BS communicates to all users simultaneously through an error free link of capacity $C$ bits per unit time. In this case, the coded multicast codeword consists simply of the concatenation of the coded  messages
\begin{equation} \nonumber
X = \{ U_S : S \subseteq [1:K], |S| = t + 1\}. 
\end{equation}
The transmission length of $X$ is 
\begin{equation} \nonumber
\sum_S \mathcal{L}(U_S) F = \frac{K (1 - M/N)}{1 + MK/N} F, 
\end{equation}
 resulting in the symmetric rate 
\begin{equation}
 R_{\mathrm{sym}} = \frac{C}{\sum_S \mathcal{L}(U_S)} = \frac{C (1 + MK/N)}{K (1 - M/N)} 
\end{equation} 
(consistently with our definition given before). 

Now, it turns out that  each coded message $U_S$ is useful only to the users in $S$. Therefore, each message $U_S$ can be sent by multicasting to the group of users $S$. In order to do this we use a beamforming vector $\mathbf{w}_S$ (where $||\mathbf{w}_S|| \leq 1$) which results in the following common rate for group $S$:
\begin{equation}\label{Eq_Groupcast_Rate}
\min_{k \in S}\log \left(1+|\mathbf{h}_k^H \mathbf{w}_S|^2 SNR\right),
\end{equation}
which can be maximized by choosing
\begin{eqnarray}\label{Eq_MaxMinOpt}
\mathbf{w}_S^*&=& \arg \max_{\mathbf{w}} \min_{k \in S} |\mathbf{h}_k^H \mathbf{w}|, \\ \nonumber
\mathrm{s.t.} && ||\mathbf{w}||^2 \leq 1.
\end{eqnarray}
This optimization problem has been shown to be NP-Hard, however, close-to-optimal solutions can be obtained by a Semidefinite Relaxation (SDR) approach \cite{Sidiropoulos_2016}.

This will result in the symmetric rate
\begin{equation} \label{Eq_MaxMin_SymRate}
R_{\mathrm{sym}}^{(1)} = \frac{F}{T} = \left [  \sum_{\substack{S \subset [1:K] \\ |S|=t+1}} \frac{1/ {K \choose t}}{\log \left(1+\min_{k \in S}|\mathbf{h}_k^H \mathbf{w}_S^*|^2 SNR\right)} \right ]^{-1}
\end{equation}
where $\mathbf{w}^*_S$ is the solution to (\ref{Eq_MaxMinOpt}).

\section{Multi-Antenna Coded Caching, \\ Linear Combination in the Complex Field}\label{Sec:ComplexField}

The basic idea of the last section was to adapt the multicasting opportunities  to the wireless channel via max-min fair beamforming. However, in the above scheme the spatial multiplexing gain of transmit antennas is not exploited. In this section we introduce a new scheme which exploits multiplexing and caching gains simultaneously. This scheme borrows the idea of simultaneous Zero-Forcing and Coded Caching in the delivery phase from \cite{Pooya_2016}, which is here adapted to the wireless scenario. Let us first explain the main idea through an example.

\begin{algorithm}[t]
\caption{Multi-Antenna Coded Caching - Complex Field Subfile Combination \label{Alg_CC_Complex}}
\begin{algorithmic}[1]

\Procedure{DELIVERY}{$W_1,\dots,W_N$, $d_1,\dots,d_K$, $H$}
\State $t \gets MK/N$

\State INDEX-INIT

\ForAll{$S \subseteq [K], |S|=t+L$}
\ForAll{$T \subseteq S, |T|=t+1$}

$\quad \quad\quad\mathbf{u}_S^T=$BFV($S$, $T$,$H$)
\EndFor

\ForAll {$\omega=1,\dots,{t+L-1 \choose t}$}
\ForAll{$T \subseteq S, |T|=t+1$}
\State $G_\omega(T) \gets L_{r \in T}^\omega \left( \tilde{W}_{{d_r},T\backslash\{r\}}^{N(r,T\backslash\{r\})} \right)$
\EndFor
\State $\mathbf{X}_\omega(S) \gets \sum_{T \subseteq S, |T|=t+1} {\mathbf{u}_{S}^{T} G_\omega(T)}$
\EndFor

\State \textbf{transmit} $\mathbf{X}(S)=\left[\mathbf{X}_1(S),\dots,\mathbf{X}_{{t+L-1 \choose t}}(S)\right]$ with rate in \eqref{Eq_SubsetRate_Complex}.

\State INDEX-UPDATE

\EndFor
\EndProcedure

\end{algorithmic}
\end{algorithm}

\begin{algorithm}[t]
\caption*{Auxiliary Procedures.\label{Alg_CC_Aux}}
\begin{algorithmic}[1]

\Procedure{$\mathbf{u}_S^T=$BFV}{$S$, $T$,$H$}
\State Design $\mathbf{u}_{S}^{T}$ such that: for all $j \in S$, $\mathbf{h}_j \perp  \mathbf{u}_{S}^{T}$ if $j \not\in T$ and $\mathbf{h}_j \not \perp  \mathbf{u}_{S}^{T}$ if $j \in T$, and $||\mathbf{u}^T_S||=1$
\EndProcedure

-------------------------
\Procedure{Index-Init}{}
\ForAll{$T \subseteq [K], |T|=t+1$}
\ForAll{$r \in T$}
\State $N(r,T\backslash\{r\}) \gets 1 $
\EndFor
\EndFor
\EndProcedure

-------------------------
\Procedure{Index-Update}{}
\ForAll{$T \subseteq S, |T|=t+1$}
\ForAll{$r \in T$}
\State $N(r,T\backslash\{r\}) \gets N(r,T\backslash\{r\}) + 1$
\EndFor
\EndFor
\EndProcedure

\end{algorithmic}
\end{algorithm}

\begin{examp}\label{Examp_Complex}
Here we consider $L=2$ transmit antennas, $K=3$ users, $N=3$ files  $\{W_1,W_2,W_3\}=\{A,B,C\}$, and $M=1$. In the first phase, each file is divided into three equal-sized parts and in the cache content placement phase the caches are filled as:
\begin{eqnarray} \nonumber
Z_1=\{A_1,B_1,C_1\}, 
Z_2=\{A_2,B_2,C_2\},  
Z_3=\{A_3,B_3,C_3\}.
\end{eqnarray}
Then, the transmitter will send the following blocks sequentially \small
\begin{align}
\underline{\mathbf{X}}_1&=  \frac{1}{\sqrt{6}}\left[\left(
  \tilde{B}_1 +  \tilde{A}_2 \right)\frac{\mathbf{h}_3^{\perp}}{|\mathbf{h}_3^{\perp}|} 
 +\left(
   \tilde{B}_3 +  \tilde{C}_2 \right) \frac{\mathbf{h}_1^{\perp}}{|\mathbf{h}_1^{\perp}|}  \nonumber
 +\left( \tilde{A_3} +  \tilde{C_1} \right) \frac{\mathbf{h}_2^{\perp}}{|\mathbf{h}_2^{\perp}|}\right]
 \\ \nonumber
\underline{\mathbf{X}}_2&=  \frac{1}{\sqrt{6}}\left[\left(
  \tilde{B}_1 +  \tilde{A}_2 \right)\frac{\mathbf{h}_3^{\perp}}{|\mathbf{h}_3^{\perp}|} 
 +\left( \tilde{C}_2
   -\tilde{B}_3    \right) \frac{\mathbf{h}_1^{\perp}}{|\mathbf{h}_1^{\perp}|}  \nonumber
 -\left( \tilde{A_3} +  \tilde{C_1} \right) \frac{\mathbf{h}_2^{\perp}}{|\mathbf{h}_2^{\perp}|}\right]
\end{align} \normalsize
where $\underline{\mathbf{X}}_1$ and $\underline{\mathbf{X}}_2$ are $L \times \frac{Fn}{3m}$ complex blocks, requiring $\frac{Fn}{3m}$ channel uses. The first user will receive $\mathbf{h}_1^H \underline{\mathbf{X}}_1+\mathbf{z}_1^{(1)}$ and $\mathbf{h}_1^H \underline{\mathbf{X}}_2+\mathbf{z}_1^{(2)}$ sequentially, and with the help of its cache contents, the first user extracts
\begin{equation} \nonumber
\frac{1}{\sqrt{3}} \mathbf{U} \left( \begin{array}{cc}
\frac{\mathbf{h}_1^H \mathbf{h}_3^{\perp}}{|\mathbf{h}_3^{\perp}|}  & 0\\
0 &  \frac{\mathbf{h}_1^H \mathbf{h}_2^{\perp}}{|\mathbf{h}_2^{\perp}|} \end{array} \right)  \left( \begin{array}{c}
\tilde{A}_2   \\
\tilde{A}_3    \end{array} \right) + \left( \begin{array}{c}
\mathbf{\underline{z}}_1^{(1)}   \\
\mathbf{\underline{z}}_1 ^{(2)}   \end{array} \right),
\end{equation}
where $\mathbf{U}$ is the unitary matrix
\begin{equation} \nonumber
\mathbf{U} = \frac{1}{\sqrt{2}}\left( \begin{array}{cc}
1  & 1\\
1 &  -1 \end{array} \right).
\end{equation}
This user multiplies its received signal by $\mathbf{U}^H$, and then can decode its desired messages if encoding rate of subfiles $A_2$ and $A_3$ is less than:
\begin{equation}\nonumber
\log \left(1+\frac{1}{3}\min \left(\frac{|\mathbf{h}_1^H \mathbf{h}_2^{\perp}|^2}{|\mathbf{h}_2^{\perp}|^2},\frac{|\mathbf{h}_1^H \mathbf{h}_3^{\perp}|^2}{|\mathbf{h}_3^{\perp}|^2}\right) SNR \right).
\end{equation}
Thus, considering the minimum rate of all the three users, the common transmission rate to all three users should be less than
\begin{equation} \nonumber
\log \left(1+\frac{1}{3}\min_{\substack{i,j \in\{1,2,3\} \\ i \neq j}} \left(\frac{|\mathbf{h}_i^H \mathbf{h}_j^{\perp}|^2}{|\mathbf{h}_j^{\perp}|^2}\right) SNR \right).
\end{equation}
This results in the symmetric rate:

\begin{eqnarray}\label{Eq_Ex1_SymRate}
 R_{\mathrm{sym}} 
= \frac{3}{2} \log \left(1+\frac{1}{3}\min_{\substack{i,j \in\{1,2,3\} \\ i \neq j}} \left(\frac{|\mathbf{h}_i^H \mathbf{h}_j^{\perp}|^2}{|\mathbf{h}_j^{\perp}|^2}\right) SNR \right).
\end{eqnarray}

\end{examp}

The scheme in Example \ref{Examp_Complex} can be extended to the general case of $K$, $N$, $L$, and $M$, following the same guidelines in \cite{Pooya_2016}, adapted to the wireless scenario. This generalization is shown in Algorithm \ref{Alg_CC_Complex}. Theorem \ref{Th_SymRate_Complex} characterizes the symmetric rate of this algorithm.

\begin{thm}\label{Th_SymRate_Complex}
Algorithm \ref{Alg_CC_Complex} will result in the following symmetric rate
\begin{eqnarray}\label{Eq_Sym_Rate_Complex} \nonumber
R_{\mathrm{sym}}^{(2)}=\frac{{K \choose t}{K-t-1 \choose L-1}}{{t+L-1 \choose t}} 
\left[\sum_{\substack{S \subseteq [1:K] \\ |S|=t+L}} \left(R_C^{(2)}(S)\right)^{-1}\right]^{-1}, \\
\end{eqnarray}
where \small
\begin{eqnarray}\label{Eq_SubsetRate_Complex} 
R_C^{(2)}(S)= \log \left(1+\frac{SNR}{t+L}  \min_{\substack{T \subseteq S \\ |T|=t+1}} \min_{r\in T} |\mathbf{h}_r^H \mathbf{u}^{T}_S|^2 \right).
\end{eqnarray} \normalsize
\end{thm}
\begin{proof}
Please refer to the Appendix A.
\end{proof}

\section{Multi-Antenna Coded Caching, \\ Linear Combination in the Finite Field} \label{Sec:FiniteField}

\begin{algorithm}[t]
\caption{Multi-Antenna Coded Caching - Finite Field Subfile Combination\label{Alg_CC_Finite}}
\begin{algorithmic}[1]

\Procedure{DELIVERY}{$W_1,\dots,W_N$, $d_1,\dots,d_K$, $H$}
\State $t \gets MK/N$
\State INDEX-INIT
\ForAll{$S \subseteq [K], |S|=t+L$}
\ForAll{$T \subseteq S, |T|=t+1$}
\State $\mathbf{u}_S^T=$BFV($S$, $T$,$H$)

\EndFor

\ForAll{$T \subseteq S, |T|=t+1$}
\State $G'(T) \gets \oplus_{r \in T}  W_{{d_r},T\backslash\{r\}}^{N(r,T\backslash\{r\})} $
\EndFor
\State $\mathbf{X}(S) \gets \sum_{T \subseteq S, |T|=t+1} {\frac{\mathbf{u}_{S}^{T}}{\sqrt{t+L \choose t+1}} \psi\left(G'(T)\right)}$

\State \textbf{transmit} $\mathbf{X}(S)$ with rate in \eqref{Eq_SubsetRate_Finite}.

\State INDEX-UPDATE

\EndFor
\EndProcedure

\end{algorithmic}
\end{algorithm}

In this section we analyze a similar scheme, but here the coded caching messages are developed in the finite field, and then modulated to complex numbers. Let us first revisit our example:

\begin{examp}\label{Examp_Finite}
Here the problem setup and the cache content placement is the same as Example 1, but the second phase is different.
In the delivery phase, the transmitter will send $\underline{\mathbf{X}}=$ \small
\begin{align} \nonumber
\frac{1}{\sqrt{3}} \left[
  \psi \left(B_1 \oplus  A_2\right) \frac{\mathbf{h}_3^{\perp}}{|\mathbf{h}_3^{\perp}|}  
 +\psi\left(
   B_3 \oplus  C_2 \right) \frac{\mathbf{h}_1^{\perp}}{|\mathbf{h}_1^{\perp}|}   
 +\psi\left( A_3 \oplus  C_1 \right) \frac{\mathbf{h}_2^{\perp}}{|\mathbf{h}_2^{\perp}|}   \right]
 \end{align} \normalsize
where $\underline{\mathbf{X}}$ is a $L \times \frac{Fn}{3m}$ complex block, requiring $\frac{Fn}{3m}$ channel uses. Let us focus on the first user who will receive $\mathbf{h}_1^H \underline{\mathbf{X}}+\mathbf{z}_1$ as follows
\begin{align}\nonumber
\mathbf{h}_1^H \underline{\mathbf{X}}+\mathbf{z}_1 =  \psi(B_1 \oplus A_2) \frac{\mathbf{h}_1^H\mathbf{h}_3^\perp}{\sqrt{3}|\mathbf{h}_3^\perp|}  
+ \psi(A_3 \oplus C_1)\frac{\mathbf{h}_1^H\mathbf{h}_2^\perp}{\sqrt{3}|\mathbf{h}_2^\perp|}  +   \mathbf{z}_1.
\end{align}
Since User 1 is interested in decoding both $\psi(B_1 \oplus A_2)$ and $\psi(A_3 \oplus C_1)$, and their encoding is done independently, this can be considered as a MAC channel. Let us define
\begin{eqnarray}\label{Eq_Examp2_MAC_Rates} \nonumber
R_{\mathrm{Sum}}^{(1)}&=& \log\left(1+\frac{1}{3}\left(\frac{|\mathbf{h}_1^H\mathbf{h}_3^\perp|^2}{|\mathbf{h}_3^\perp|^2}+\frac{|\mathbf{h}_1^H\mathbf{h}_2^\perp|^2}{|\mathbf{h}_2^\perp|^2}\right)SNR\right), \\ \nonumber
R^{(1)}_{\{1,2\}}&=& \log\left(1+\frac{1}{3}\left(\frac{|\mathbf{h}_1^H\mathbf{h}_3^\perp|^2}{|\mathbf{h}_3^\perp|^2}\right)SNR\right), \\ 
R^{(1)}_{\{1,3\}}&=&  \log\left(1+\frac{1}{3}\left(\frac{|\mathbf{h}_1^H\mathbf{h}_2^\perp|^2}{|\mathbf{h}_2^\perp|^2}\right)SNR\right).
\end{eqnarray}
Then, if we operate the MAC channel at an equal rate point, user 1 will receive useful information with the effective sum rate:
\begin{eqnarray}\label{Eq_Examp2_Effective_Rates}
R_{\mathrm{Eff}}^{(1)}=\min\left(R_{\mathrm{Sum}}^{(1)}, 2R^{(1)}_{\{1,2\}}, 2R^{(1)}_{\{1,3\}} \right).
\end{eqnarray}
Similarly, effective sum rate for other users will be $R_{\mathrm{Eff}}^{(2)}$ and $R_{\mathrm{Eff}}^{(3)}$. Then, the symmetric rate will be:
\begin{equation}\label{Eq_Ex2_SymRate}
R_{\mathrm{sym}}=\frac{3}{2} \min\left(R_{\mathrm{Eff}}^{(1)},R_{\mathrm{Eff}}^{(2)},R_{\mathrm{Eff}}^{(3)}\right).
\end{equation}
\end{examp}
The generalization of Example 2 to general $K$, $N$, $L$, and $M$ is shown in Algorithm \ref{Alg_CC_Finite}. Theorem \ref{Th_SymRate_Finite} characterizes the symmetric rate of Algorithm \ref{Alg_CC_Finite}.
\begin{thm}\label{Th_SymRate_Finite}
Algorithm \ref{Alg_CC_Finite} will result in the following symmetric rate 
\begin{eqnarray}\label{Eq_Sym_Rate_Finite} \nonumber
R_{\mathrm{sym}}^{(3)}=\frac{{K \choose t}{K-t-1 \choose L-1}}{{t+L-1 \choose t}} 
\left[\sum_{\substack{S \subseteq [1:K] \\ |S|=t+L}} \left(R_C^{(3)}(S)\right)^{-1}\right]^{-1}, \\
\end{eqnarray} 
where \small
\begin{eqnarray}\label{Eq_SubsetRate_Finite} 
\nonumber R_C^{(3)}(S)
&=& \min _{r \in S}\min \Big[\log\Big(1+\frac{1}{{t+L\choose t+1}}\sum_{\substack{T \subseteq S \\ |T|=t+1 \\ r\in T}}|\mathbf{h}_r^H\mathbf{u}_S^T|^2SNR\Big), \\ \nonumber
&& {t+L-1 \choose t} \min_{\substack{T \subseteq S \\ |T|=t+1 \\ r\in T}} \log\Big(1+\frac{1}{{t+L\choose t+1}}|\mathbf{h}_r^H\mathbf{u}_S^T|^2SNR\Big) \Big]. \\
\end{eqnarray} \normalsize
\end{thm}
\begin{proof}
Please refer to the Appendix B.
\end{proof}

\begin{myremark}
It should be noted that the main difference between Algorithm \ref{Alg_CC_Complex} and \ref{Alg_CC_Finite} is that $G(T)$ in Algorithm \ref{Alg_CC_Complex} is a linear combination of subfiles in the complex field, while $G'(T)$ in Algorithm \ref{Alg_CC_Finite} is linear combination of subfiles in the finite field.
\end{myremark}

\section{Performance Comparison}\label{Sec:Compare}

All the above schemes exploit multiple antennas at the transmitter, in addition to multicasting opportunities provided by coded caching. It can easily be verified that by defining per user $DoF_{i}=\lim_{SNR \rightarrow \infty}(R_{\mathrm{sym}}^{(i)}/ \log SNR)$ we will have
\begin{align}\label{Eq_DoF_Compare}\nonumber
DoF_1&=\frac{1+KM/N}{K(1-M/N)}, \\ 
DoF_2&=DoF_3=\frac{L+KM/N}{K(1-M/N)}, 
\end{align}
which is consistent with the results in \cite{Maddah-Ali_Fundamental_2014} and \cite{Pooya_2016}. 

\begin{figure}[t]
\begin{center}
\includegraphics[width=0.5\textwidth]{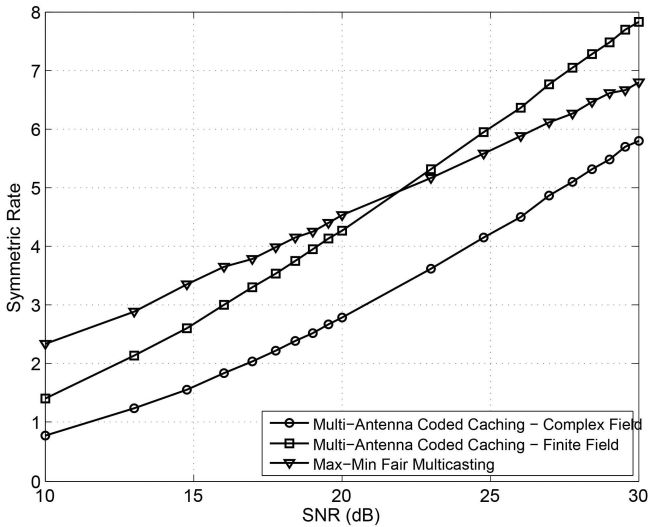}
\end{center}
\caption{Low and Intermediate SNR Rate Comparison. Here we have $K=3$, $L=2$, and $t=1$. \label{Fig_Low_SNR}}
\end{figure}

\begin{figure}[t]
\begin{center}
\includegraphics[width=0.5\textwidth]{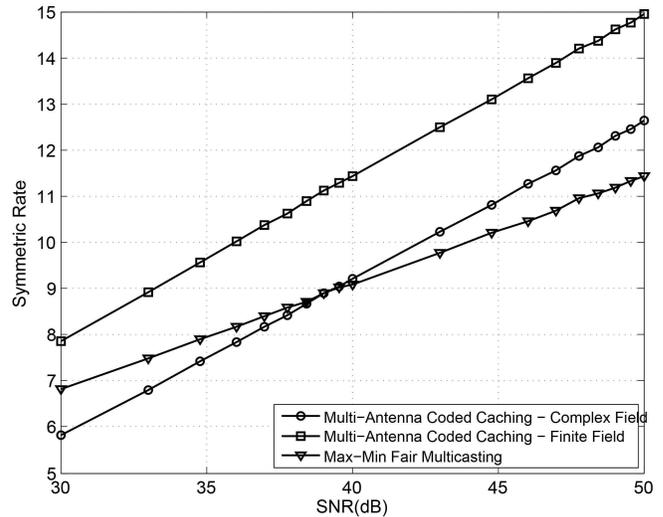}
\end{center}
\caption{High SNR Rate Comparison. Here we have $K=3$, $L=2$, and $t=1$. \label{Fig_High_SNR}}
\end{figure}

For numerical illustrations at finite $SNR$ we consider the settings in Examples \ref{Examp_Complex} and \ref{Examp_Finite} in a Rayleigh fading wireless setup.
Figure \ref{Fig_Low_SNR} shows the symmetric rate of the three schemes for the range $SNR = [10dB\sim30dB]$. This figure shows that at low $SNR$ the max-min fair multicasting has the best performance, while for $SNR>21dB$ the multi-antenna coded caching with finite field combination has the best performance. Figure \ref{Fig_High_SNR} shows symmetric rate for the range $SNR = [30dB\sim50dB]$. In this figure we see that at high $SNR$ both multi-antenna coded caching schemes (complex and finite field linear combinations) surpass the max-min fair approach, which is consistent with the DoF analysis above.

\section{Conclusions}\label{Sec:Conclusions}

In this paper we have considered a cache-enabled MISO broadcast scenario, and have investigated the performance of a scheme benefiting from multiplexing and caching gains at the same time. We have derived finite SNR rate expressions for the proposed scheme. Although the proposed scheme is shown to surpass a max-min fair multicast baseline scheme in terms of DoF performance, careful modifications are required for proper performance at low SNR values. .

\clearpage 
\appendices
%\section{}

\section{}
\begin{proof}[Proof of Theorem \ref{Th_SymRate_Complex}]

For the general  $K$, $L$, $N$, and $M$ case, the placement procedure is the same as \cite{Maddah-Ali_Fundamental_2014}. However, here we divide each sub-packet further into ${K-t-1 \choose L-1}$ mini-packets of equal size as follows:
\begin{equation} \nonumber
W_{n,\tau} = \left\{W_{n,\tau}^j: j=1,\dots,{K-t-1 \choose L-1}\right\}. 
\end{equation}
The delivery phase is based on the ideas proposed in \cite{Pooya_2016}, adapted to the wireless medium, which is represented in Algorithm \ref{Alg_CC_Complex}. 

Consider a $(t+L)$-subset of users named $S$. This subset has $q={t+L \choose t+1}$ number of $(t+1)$-subsets which we call $T_i, i=1,\dots,q$. Our aim is to deliver
\begin{eqnarray}\label{Eq_Gw_def} \nonumber
G_{\omega}(T_i)&=& L_{r \in T_i}^\omega \left( \tilde{W}_{{d_r},T_i\backslash\{r\}} \right) \\ 
&=&\sum_{r \in T_i}{\sigma^\omega_{{d_r},T_i\backslash\{r\}}  \tilde{W}_{{d_r},T_i\backslash\{r\}} },
\end{eqnarray}
to the subset $T_i$. Here $L_{r \in T_i}^\omega$ means a random linear combination. Upon successful reception of this linear combination at $T_i$, each user $r \in T_i$ can, with the help of its cache contents,  decode $W_{{d_r},T_i\backslash\{r\}}$.

The main idea in Algorithm \ref{Alg_CC_Complex} is that with the help of the Zero-Forcing techniques we can deliver $G_{\omega}(T_i)$ to all $T_i \subseteq S$ simultaneously. The main challenge is that the members of $S \backslash T_i$ are not interested in receiving $G_{\omega}(T_i)$. Thus, to avoid interference, $G_{\omega}(T_i)$ is zero-forced at these $L-1$ users. For doing the same idea for all $T_i \subseteq S$ the transmitter should send
\begin{equation}\label{Eq_X_omega_S}
\mathbf{X}_\omega(S) = \sum_{T \subseteq S, |T|=t+1} {\mathbf{u}_{S}^{T} G_\omega(T)},
\end{equation}
where we design $\mathbf{u}_{S}^{T}$ such that: for all $j \in S$, $\mathbf{h}_j \perp  \mathbf{u}_{S}^{T}$ if $j \not\in T$ and $\mathbf{h}_j \not \perp  \mathbf{u}_{S}^{T}$ if $j \in T$, and $|\mathbf{u}_{S}^{T}|=1$. This task is done by the procedure BFV in Algorithm \ref{Alg_CC_Complex}, where its details are shown in Auxiliary Procedures.

The index $\omega$ in $L_{r \in T}^\omega$ ensures that, with appropriate choice of coefficients, each user will receive independent linear combinations of its desired mini-files. The index  $N(.,.)$ in Algorithm \ref{Alg_CC_Complex} ensures that each user will not receive the same sub-file more than once. The Procedures \emph{INDEX-INIT} and \emph{INDEX-UPDATE} will take care of that concern in Algorithm \ref{Alg_CC_Complex}, where their details are shown in Auxiliary Procedures. 

It has been shown in \cite{Pooya_2016} that, in a \emph{noise-free} scenario the Algorithm \ref{Alg_CC_Complex} will succeed in delivering the requested files to all the users. However, in the current finite SNR wireless scenario we should limit the transmit rate of all the mini-files encoders to guarantee successful delivery at all the receivers. The following lemma characterizes the maximum transmission rate to each subset of users, at which all these users can decode their required mini-files successfully.

\begin{lem}\label{Lem_RCS_Complex}
Consider a $(t+L)$-subset of users $S$. Then, if the common transmission rate to this subset according to Algorithm \ref{Alg_CC_Complex} is less than
\begin{eqnarray}\label{Eq_Lemma1} \nonumber
R_C(S)&=& \log \left(1+\frac{SNR}{t+L}  \min_{r \in S} \min_{\substack{T \subseteq S \\ r\in T}} |\mathbf{h}_r^H \mathbf{u}^{T}_S|^2 \right) \\ 
&=& \log \left(1+\frac{SNR}{t+L}  \min_{T \subseteq S} \min_{r\in T} |\mathbf{h}_r^H \mathbf{u}^{T}_S|^2 \right),
\end{eqnarray}
then all the users can decode their required mini-files successfully.
\end{lem}

In order to prove Lemma \ref{Lem_RCS_Complex} let us consider user $r \in S$ in a fixed $(t+L)$-subset $S$. This user will receive 
\begin{eqnarray}\label{Eq_Lem1Proof_Main} \nonumber
\mathbf{h}_r^H \mathbf{X}_{\omega}(S)&=&\sum_{\substack{T \subseteq S \\ |T|=t+1}} \mathbf{h}_r^H\mathbf{u}_{S}^{T} G_{\omega}(T) \\ \nonumber
&\stackrel{(a)}=& \sum_{\substack{T \subseteq S \\ |T|=t+1}} \mathbf{h}_r^H{\mathbf{u}_{S}^{T} \sum_{i \in T}{\sigma^\omega_{{d_i},T\backslash\{i\}}   \tilde{W}_{{d_i},T\backslash\{i\}} }} \\ \nonumber
&\stackrel{(b)}=&\sum_{\substack{T \subseteq S \\ |T|=t+1 \\ r \in T  }} \mathbf{h}_r^H{\mathbf{u}_{S}^{T} \sum_{i \in T}{\sigma^\omega_{{d_i},T\backslash\{i\}}   \tilde{W}_{{d_i},T\backslash\{i\}} }} \\ 
&\stackrel{(c)}=& \sum_{\substack{T \subseteq S \\ |T|=t+1 \\ r \in T  }}  \mathbf{h}_r^H{\mathbf{u}_{S}^{T} \sigma^\omega_{{d_r},T\backslash\{r\}}   \tilde{W}_{{d_r},T\backslash\{r\}} }, 
\end{eqnarray}
for all $\omega=1,\dots,{t+L-1 \choose t}$. 
In \eqref{Eq_Lem1Proof_Main},  $(a)$ follows from \eqref{Eq_Gw_def}, $(b)$ follows from the fact that $\mathbf{h}_r^H\mathbf{u}_S^T=0$ if $r \not\in T$, and $(c)$ follows from the fact that this user has cached, and thus can eliminate $\tilde{W}_{d_i,T \backslash \{i\}}$ for all $i \neq r$.

Let us define 
\begin{equation}\nonumber
\mathbf{W}^{S}_{r}=\left( \begin{array}{c}
  \tilde{W}_{{d_r},T_1\backslash\{r\}}   \\ \vdots \\  \tilde{W}_{{d_r},T_v\backslash\{r\}}   \end{array} \right),
\end{equation}
 and the $v \times v$ matrix $\mathbf{L}_r^S$ such that
\begin{eqnarray}\nonumber
\mathbf{L}_r^S(\omega,i)= \mathbf{h}_r^H\mathbf{u}_{S}^{T_i} \sigma^\omega_{{d_r},T_i\backslash\{r\}}
\end{eqnarray}
for all $i,\omega=1,\dots,{t+L-1 \choose t}$, where $T_1, \dots T_v \in S, r \in T_i$ for $v={t+L-1 \choose t}$.
Then, the observations of this user can be shown as
\begin{equation}\nonumber
\mathbf{L}_r^S \mathbf{W}^{S}_{r}, +\mathbf{z}_r,
\end{equation}
where it is easy to verify that the coefficients $\sigma$ can be chosen such that we have:
\begin{equation}\label{Eq_Decomposition}
\mathbf{L}_r^S=\sqrt{\frac{{t+L-1 \choose t}}{(t+1){t+L \choose t+1}}} \mathbf{U}  \left( \begin{array}{ccc}
\mathbf{h}_r^H \mathbf{u}^{T_1}_S & \dots & 0 \\ \vdots & \ddots & \vdots \\ 0  &\dots& 0 \\ 0 & \dots & \mathbf{h}_r^H \mathbf{u}^{T_v}_S \end{array} \right),
\end{equation}
where $\mathbf{U}$ is a ${t+L-1 \choose t} \times {t+L-1 \choose t}$ unitary matrix. Also we have used the fact that if all the coefficients satisfy
\begin{equation}\nonumber
|\sigma|^2 \leq \frac{1}{ {t+L \choose t+1}(t+1)},
\end{equation}
then the transmit power constraint in \eqref{Eq_Transmit_Power_Const} is satisfied.

Thus, user $r$ can decode all its required data if the transmission rate is less than
\begin{equation} \nonumber
\log \left(1+\frac{SNR}{t+L}   \min_{\substack{T \subseteq S \\ r\in T \\ |T|=t+1}} |\mathbf{h}_r^H \mathbf{u}^{T}_S|^2 \right).
\end{equation}
Considering the fact that we need all the users in $S$ to successfully decode, they should have the common rate
\begin{equation}\label{Eq_Th1_Proof_Subset_Rate}
\log \left(1+\frac{SNR}{t+L}   \min_{r \in S} \min_{\substack{T \subseteq S \\ r\in T \\|T|=t+1}} |\mathbf{h}_r^H \mathbf{u}^{T}_S|^2 \right).
\end{equation}

It should be noted that, based on Lemma \ref{Lem_RCS_Complex}, each user in $S$ can receive data with rate \eqref{Eq_Th1_Proof_Subset_Rate}. Since each user in $S$ decodes 
\begin{equation}\nonumber
\frac{{t+L-1 \choose t}}{{K \choose t}{K-t-1 \choose L-1}}F
\end{equation}
bits after transmission to $S$ is concluded, the symmetric rate will be equal to \eqref{Eq_Sym_Rate_Complex}, and the proof is complete.

\end{proof}

\section{}
\begin{proof}[Proof of Theorem \ref{Th_SymRate_Finite}]
Algorithm \ref{Alg_CC_Finite} is the same as Algorithm \ref{Alg_CC_Complex} in the cache content placement phase. The only difference is in the delivery phase. In Algorithm \ref{Alg_CC_Complex} the coded chunks are formed in the signal domain (complex field), while in Algorithm \ref{Alg_CC_Finite} they are formed in the data domain (finite field). Thus, let us first form the coded chunks as
\begin{eqnarray}\label{Eq_GPrime_def}
G'(T_i)=\oplus_{r \in T_i} W_{d_r, T_i \backslash \{r\}}.
\end{eqnarray}
Then, in Algorithm \ref{Alg_CC_Finite} the transmitter will send
\begin{equation}\label{Eq_X_Transmit_Finite}
\mathbf{X} = \sum_{\substack{T \subseteq S \\ |T|=t+1}} \mathbf{u}_S^T \frac{1}{\sqrt{{t+L \choose t+1}}} \psi\left(G'(T)\right).
\end{equation}
It can be easily checked that the transmit signal in \eqref{Eq_X_Transmit_Finite} satisfies the transmit power constraint in \eqref{Eq_Transmit_Power_Const}.Then, the following lemma characterizes the rate for successful transmission of $\mathbf{X}$.
\begin{lem}\label{Lem_RCS_Finite}
Consider a $(t+L)$-subset of users $S$. 
Then, if the common transmission rate to this subset according to Algorithm \ref{Alg_CC_Finite} is less than
\begin{eqnarray}\label{Eq_Lemma2}
R_C(S)=\min_{r \in S} R_{\mathrm{Eff}}^{(r)},
\end{eqnarray}
where \small
\begin{eqnarray} \nonumber
R_{\mathrm{Eff}}^{(r)}&=& \min \Big[\log\Big(1+\frac{1}{{t+L\choose t+1}}\sum_{\substack{T \subseteq S \\ |T|=t+1 \\ r\in T}}|\mathbf{h}_r^H\mathbf{u}_S^T|^2SNR\Big), \\ \nonumber
&& {t+L-1 \choose t} \min_{\substack{T \subseteq S \\ |T|=t+1 \\ r\in T}} \log\Big(1+\frac{1}{{t+L\choose t+1}}|\mathbf{h}_r^H\mathbf{u}_S^T|^2SNR\Big) \Big],
\end{eqnarray} \normalsize
this transmission will be successful.
\end{lem}

In order to prove Lemma \ref{Lem_RCS_Finite}, let us focus on user $r$ who will receive
\begin{eqnarray}\nonumber
\mathbf{h}_r^H\mathbf{X}(S) +\mathbf{z}_r &=& \sum_{\substack{T \subseteq S \\ |T|=t+1}}  \frac{\mathbf{h}_r^H\mathbf{u}_S^T}{\sqrt{{t+L \choose t+1}}} \psi\left(G'(T)\right) +\mathbf{z}_r \\ \nonumber
&=& \sum_{\substack{T \subseteq S \\ |T|=t+1 \\ r \in T}}  \frac{\mathbf{h}_r^H\mathbf{u}_S^T}{\sqrt{{t+L \choose t+1}}} \psi\left(G'(T)\right) +\mathbf{z}_r.
\end{eqnarray}

Considering the fact that User $r$ is interested in decoding all the ${t+L-1 \choose t}$ terms in above summation with equal rates, and using achievable capacity region of MAC channels \cite{Cover}, the lemma proof is complete.

Now, since each user in $S$ decodes 
\begin{equation}\nonumber
\frac{{t+L-1 \choose t}}{{K \choose t}{K-t-1 \choose L-1}}F
\end{equation}
bits after transmission to $S$ is concluded, the symmetric rate will be equal to \eqref{Eq_Sym_Rate_Finite}, and the proof of Theorem 2 is complete.

\end{proof}

%\section*{Acknowledgment}
%...

\end{document}